\documentclass[nonatbib, nocopyrightspace]{sigplanconf}
\usepackage[utf8]{inputenc}
\usepackage[english]{babel}

\usepackage[hang,flushmargin]{footmisc}
\usepackage{amsmath}
\usepackage{stmaryrd}
\usepackage[frozencache]{minted}
\usepackage{multicol}
\usepackage{verbatim}
\usepackage{float}
\usepackage{todonotes}
\usepackage{wrapfig, placeins}
\usepackage[acronym,shortcuts]{glossaries}
\usepackage[hyphens]{url}
\usepackage{hyperref}
\hypersetup{breaklinks=true}
\usepackage{cleveref}
\usepackage{url}

\makeatletter
\providecommand{\@thefnmark}{\the\@fnmark}
\makeatother

\AtBeginEnvironment{minted}{%
  }

\newcommand{\h}[1]{\mintinline{haskell}{#1}}
\newcommand{\hnh}[1]{\mintinline{text}{#1}} 
\newcommand{\hc}[1]{\begin{center}\mintinline{haskell}{#1}\end{center}}

\makeglossaries
\newacronym{GADT}{GADT}{Generalized Algebraic Data Type}
\newacronym{SMT}{SMT}{Satisfiability Modulo Theories}
\newacronym{GHC}{GHC}{Glasgow Haskell Compiler}
\newacronym{AST}{AST}{Abstract Syntax Tree}

\begin{document}
\preprintfooter{short description of paper}   

\title{Structural and semantic pattern matching analysis in Haskell}
\subtitle{Analysis of exhaustiveness, redundancy, and divergence of pattern matching\titlenote{
This report has been originally written in the summer of 2016. Due to time constraints, it has not been published at the time. The authors have decided to make it public ex post facto for posterity, but the contents should be interpreted within the contemporary frame of reference. Both authors were students at ETH Z{\"u}rich at the time of their involvement.}
}

\authorinfo{Pavel Kalvoda}{me@pavelkalvoda.com}

\authorinfo{Tom Sydney Kerckhove}{syd@cs-syd.eu}

\maketitle

\begin{abstract}
Haskell functions are defined as a series of clauses consisting of patterns that are matched against the arguments in the order of definition.
In case an input is not matched by any of the clauses, an error occurs.
Therefore it is desirable to statically prove that the function is defined for all well-typed inputs.
Conversely, a clause that can never be matched also indicates a likely defect.
Analyzing these properties is challenging due to presence of \acrfullpl{GADT} and guards as well as due to Haskell's lazy evaluation.
We implement a recently proposed algorithm that unifies and extends the related analyses implemented in version 7 of the Glasgow Haskell Compiler.
By using an \acrfull{SMT} solver to handle the semantic constraints arising from pattern guards, we achieve a further improvement in precision over the existing \acrfull{GHC} 8.0.1 implementation. 
We present a tool that uses the analysis to give sound, more precise, and actionable warnings about program defects. 
\end{abstract}

\keywords
Haskell, pattern matching, \acs{SMT}, guards, value type equality constraints.

\section{Introduction}

In Haskell, functions are defined as one or more \emph{clauses} that consist of one \emph{pattern} for each of the formal parameters. When a function application is evaluated, the program attempts to match the arguments against the clauses in the order of definition and the right hand side of the first clauses that matched is evaluated. If no clause was matched, the program terminates with an error.

For example, consider the following function:
\begin{minted}{haskell}
pairs :: [a] -> [(a, a)]
pairs [] = []
pairs (x:y:zz) = (x, y):pairs zz
\end{minted}
Calling \h{pairs} with  \h{[1]} will result in a pattern matching error and ultimately an erroneous termination. This is synonymous to the return value being \h{undefined} or \emph{bottom}, a special value that is a member of every type and indicates an unsuccessful computation.

A Haskell function that terminates with non-bottom value for all non-bottom inputs is called \emph{total}.
A necessary but not sufficient\footnote{For example, consider \h{\_ -> undefined}.} condition for totality is \emph{exhaustiveness}, the ability of the clauses to match any non-bottom input.


While there are a number of non-total functions in the standard library (e.g. \h{tail} is not defined for the empty list~\cite{partial}), the general trend is to prefer total functions \cite{avoidPartial}.
This is because a programmer should ideally not need to familiarize themselves with the implementation of a function to discover on which parts of its corange it is total before using that function.

Since totality in a Turing-complete language is trivially undecidable and a reduction in power is often impractical~\cite{DBLP:journals/jucs/ATurner04}, it is useful to examine exhaustiveness as a proxy to proving partiality, as it reveals common programming errors. It is especially useful when adding new constructors to existing types because then the compiler will notify the programmer of functions that have become partial as a consequence.

Apart from regular patterns, Haskell also supports \emph{guards}, arbitrary Boolean expressions that are evaluated when all patterns are matched. If the guard expression evaluates to \h{True}, the clause is selected, otherwise matching falls through to the next clause.

This makes checking for exhaustiveness challenging, as illustrated by the following example:

\begin{minted}[tabsize=4]{haskell}
abs :: Int -> Int
abs x
	| x < 0 = -x
	| x >= 0 = x
\end{minted}

Proving that \h{abs} is exhaustively defined cannot be achieved by structural manipulation with type definitions alone.
Semantic insight (knowledge that \hnh{x < 0 || x >= 0} holds for all \hnh{x}) is necessary.
While generally undecidable, guards tend to be simple, rendering a limited semantic analysis realistic.

Furthermore, a clause can also be unreachable because all the values that would be matched by it are matched by the preceding clauses, as demonstrated by the second clause of \h{f} in the example below.
Such clauses are called \emph{redundant} and are another likely indicator of an error. 

The lazy evaluation of Haskell means that a function may be evaluated when its result is to be matched against a pattern, even if not used later.
This leads to unexpected subtleties in the semantics of pattern matching.
For instance, consider functions \h{f} and \h{f'}:

\begin{minted}{haskell}
f :: Bool -> Bool -> Int
f _    True  = 1
f True True  = 2
f _    False = 3
\end{minted}

\begin{minted}{haskell}
f' :: Bool -> Bool -> Int
f' _ True  = 1
f' _ False = 3
\end{minted}

Although the second clause of \h{f} is clearly redundant in the sense that it is never matched, removing it changes the semantics, as demonstrated by the difference in evaluation of the following expressions:
\begin{minted}{haskell}
f undefined False
\end{minted}
\begin{minted}{haskell}
f' undefined False
\end{minted}

The evaluation of \h{f undefined False} will terminate with an error, whereas \h{f' undefined False} will evaluate to \h{3} because evaluation of the first tuple element was not forced by the removed clause. 
This is an unusual behavior that the programmer might not have introduced intentionally.

Moreover, the example illustrates that reasoning about which parts of the input will be evaluated is non-trivial, especially for recursive data types. 
Analyzing the depth of evaluation with respect to input values is thus another related topic deserving attention.

Despite their practical value, until very recently, all of the aforementioned issues were addressed only in specific cases in the \acs{GHC} \cite{marlow2004glasgow, DBLP:conf/icfp/KarachaliasSVJ15}.

\subsection{Our contributions}

In this report, we present a static analysis tool that can give accurate warnings and information for the aforementioned properties. We implement the recent algorithm by Karachalias, Schrijvers, Vytiniotis, and Jones \cite{DBLP:conf/icfp/KarachaliasSVJ15} that enables us to to overcome the challenges posed by laziness and guards while also being easily extensible to \acrshortpl{GADT}.

We simplify the existing work, provide complexity bounds, and extend it with a proof-of-concept term-constraint oracle based on an \acrshort{SMT} solver that enables semantic insight into guards at compile time. To our best knowledge, ours is the first practical implementation of this technique. We show that a similar approach is applicable to virtually all languages with similar semantics and could improve the completeness of type checking.

We also introduce the concept of \emph{evaluatedness} of a function. This is a comprehensive overview of how, when and how deeply each argument to a function will be evaluated during pattern matching.

The information our tool provides can thus be used to prevent defects, debug existing code, and help gain insight to those unfamiliar with a particular code base or Haskell in general.

\begin{figure*}[t]
	\centering
	\begin{tabular}{lllll}
		[\textsc{UNil}]&
		$\mathcal{U}$  $\epsilon$ & $\left(\Gamma, \epsilon, \Delta\right)$  & $= \left\{(\Gamma, \epsilon, \Delta)\right\}$ \vspace{0.5em}\\

		[\textsc{UConCon}]&
		$\mathcal{U}$  
		$\left(K_i \overrightarrow{p}\right)\overrightarrow{q}$ & 
		$\left(\Gamma, \left(K_j \overrightarrow{u}\right)\overrightarrow{w}, \Delta\right)$  & 
		$= \begin{cases}
		
		\texttt{map } (\texttt{kcon } K_i) \; \left(\mathcal{U} \left(\overrightarrow{p}\overrightarrow{q}\right)  \left(\Gamma,  \overrightarrow{u}\overrightarrow{w}, \Delta\right)\right)        & \text{if } K_i = K_j \\
		\left(\Gamma, \left(K_j \overrightarrow{u}\right)\overrightarrow{w}, \Delta\right) & \text{if } K_i \neq K_j\\
		\end{cases}$ \vspace{0.5em}\\

		[\textsc{UConVar}]&
		$\mathcal{U}$  
		$\left(K_i \overrightarrow{p}\right)\overrightarrow{q}$ & 
		$\left(\Gamma, x \overrightarrow{u}, \Delta\right)$  & 
		$ = \bigcup_{K_j}  \mathcal{U}   
		\left(\left(K_i \overrightarrow{p} \right) \overrightarrow{q}\right)
		(\Gamma', \left(K_j \overrightarrow{y}\right)\overrightarrow{u}, \Delta') $ \\
		
		&&&\qquad where $\overrightarrow{y}\#\Gamma$\quad
		$(x, \tau_x) \in \Gamma$ \quad
		$K_j :: \forall\overrightarrow{a}.Q \Rightarrow \overrightarrow{\tau}\rightarrow \tau$
		
		\\
		&&&	\qquad\qquad\quad					
		$\Gamma' = \Gamma \cup \left\{(\overrightarrow{y}, \overrightarrow{\tau})\right\} \cup \overrightarrow{a}$
		\\ &&&	\qquad\qquad\quad					
		$\Delta' = \Delta \cup Q \cup \left\{\tau \sim \tau_x\right\} \cup \left\{x = K_j \overrightarrow{y}\right\}$ \vspace{0.5em}\\
		
		[\textsc{UVarVar}]&																
		$\mathcal{U}$ 
		$\left(x \overrightarrow{p}\right)$ &
		$\left(\Gamma, u\overrightarrow{u}, \Delta\right) $ &
		$= \texttt{map} \left(\texttt{ucon } u \right)
		\left(
		\mathcal{U \left(\overrightarrow{p}\right)}
		\left(
		\Gamma \cup \left\{(u, \tau)\right\}
		, \overrightarrow{u}, 
		\Delta \cup \left\{x = u\right\}
		\right)
		\right)
		$ \\
		&&&\qquad where \quad $x\#\Gamma$ \quad $(u, \tau) \in \Gamma$ \vspace{0.5em}\\

		[\textsc{UGuard}]&																
		$\mathcal{U}$ 
		$\left((p, e)\overrightarrow{p}\right)$ &
		$\left(\Gamma, \overrightarrow{u}, \Delta\right) $ &
		$= \texttt{map }
		\texttt{tail}
		\left(
		\mathcal{U}
		(p \overrightarrow{p})
		\left(
		\Gamma \cup \left\{(y, \tau)\right\}
		, y\overrightarrow{u}, 
		\Delta \cup \left\{y = e\right\}
		\right)
		\right)		
		
		$
		\\
		&&&\qquad where \quad $y\#\Gamma$ \quad $(e, \tau) \in \Gamma$ \\
		\vspace{0.5em}\\

		&\texttt{kcon} $K$ &
		$\left(\Gamma, \overrightarrow{u}\overrightarrow{w}, \Delta\right)$ &
		$ = \left(\Gamma, (K\overrightarrow{u})\overrightarrow{w}, \Delta\right)$ \\
		&\texttt{ucon} $u$ &
		$\left(\Gamma, \overrightarrow{u}, \Delta\right)$ &
		$ = \left(\Gamma, u\overrightarrow{u}, \Delta\right)$ \\
		&	\texttt{tail} $\left(\Gamma, u\overrightarrow{u}, \Delta\right)$ &&
		$ = \left(\Gamma, \overrightarrow{u}, \Delta\right)$
		
	\end{tabular}
	\caption{The uncovered values function $\mathcal{U}$ and helper functions. $\epsilon$ denotes the empty vector; $Q$ is the set of existentially quantified type variables.}
	\label{tab:uncov}
\end{figure*}

\section{Background}

There exists a body of literature on the analysis of pattern matching and related problems. Initially, the problem has been examined from an efficiency perspective, since knowing the covered and uncovered values can lead to generating more specialized, performant code~\cite{Augustsson1985, DBLP:conf/icfp/FessantM01, wadler1987efficient}. Follow-up work addressed the challenge of lazy semantics~\cite{Maranget1992} as well as a limited analysis of redundancy~\cite{thiemann1993avoiding}. 

Maranget \cite{Maranget2007} introduced an algorithm for exhaustiveness and redundancy checking for the ML language, heavily borrowing from the previous compilation techniques \cite{DBLP:conf/cc/Pettersson92}. The algorithm was formulated in terms of matrices of values and includes a limited provisions for Haskell semantics and laziness while disregarding guards and \acrshortpl{GADT}.

Mitchel and Runciman \cite{Mitchell2008} gave a more sophisticated analysis for Haskell that captures all information as constraints, which enables them to precisely characterize the values, although solving these constraints has proven to be challenging.

Recently, Karachalias et al. \cite{DBLP:conf/icfp/KarachaliasSVJ15} proposed a Haskell-specific algorithm that unifies all the previous work and accounts for laziness, guards, and \acrshortpl{GADT} (see \Cref{sec:algo}). Their independent parallel work resulted in an implementation of the algorithm that became a part of \acrshort{GHC} 8 during the course of our work.

\section{Algorithm}\label{sec:algo}


In this section, we describe our adaptation of the aforementioned algorithm by Karachalias et al. \cite{DBLP:conf/icfp/KarachaliasSVJ15}. We start by providing a general intuition for the algorithm, with a precise description following in \Cref{sec:algo_out}.

\subsection{Intuition}

For the sake of simplicity, consider a well-typed program with a finite number of types and a finite number of values for each of the types. 
In such a program, we can show that a function $f$ defined using clauses~1 through $n$ of $m$ arguments each is exhaustive. 
Given a definition

\begin{equation}
\begin{aligned}
f :: \: & \tau_1 & \rightarrow \tau_2 \rightarrow \dots & \tau_m \rightarrow \tau_{m+1}\\
f\: & p_{11}&\,  p_{12} \, \dots \, & p_{1m} = \textsc{rhs}_1 \\
     &\vdots  &  & \vdots \\
f\: & p_{n1}&\, p_{n2} \, \dots \, & p_{nm} = \textsc{rhs}_n
\end{aligned}
\vspace{1em}
\end{equation}
where $\tau_1$ through $\tau_m$ are the types of the respective arguments and $\tau_{m+1}$ is the return type, it is easy to show that the function is total. 

In order to do so, we will keep track of two sets of values, $C$ and $U$, for each clause, where $C$ is intuitively the set of values that will matched by the clause and $U$ is the set of values that have not been matched by this point. Starting with the set of all well-typed inputs $U_0 = \bar{V}_{\tau_1}\times\dots \times \bar{V}_{\tau_m}$, we compute the refined sets $U_1, \dots, U_n$ of values that have not been covered yet for clauses $1$ to $n$ by just removing all value tuple covered by the respective clause: $U_k = U_{k-1} \setminus \llbracket p_{1k}\rrbracket \times \dots \times \llbracket p_{mk}\rrbracket$, where $\llbracket p\rrbracket$ is the set of values denoted by the pattern $p$, as defined by denotational semantics of Haskell~\cite[Figure 4]{DBLP:conf/icfp/KarachaliasSVJ15}.

In order to check that $f$ is exhaustive, it suffices to check that $U_n$ is empty. Furthermore, we can define the set of covered values for each clause as $C_k = U_{k-1} \cap \llbracket p_{1k}\rrbracket \times \dots \times \llbracket p_{mk}\rrbracket$. Checking whether a clause is redundant then amounts to showing that $C_k$ is empty.

While this approach is not feasible since all recursive types (e.g. lists) have infinitely many values, it can be refined by using value abstraction in the place of explicit sets of values. The presented intuition forms the basis of the algorithm proposed by Karachalias, et al. \cite{DBLP:conf/icfp/KarachaliasSVJ15}. 

Their algorithm takes advantage of the fact that all values of a given user-defined type are created using the data constructors of the type. The set of constructors provides a natural abstract domain for the set of concrete values, which in turn yields a compact representation for the sets of value tuples.

\subsection{Outline}\label{sec:algo_out}
The actual algorithm follows the structure suggested in the previous section: it processes clauses in the order of appearance, gradually refining the abstraction of values. Specifically, it computes three sets for each of the clauses:

\begin{itemize}
\item $C$, the set of \emph{covered} values. For these values, the right-hand side is evaluated.
\item $U$, the set of \emph{uncovered} values. These values will not be matched and will fall through to the next clause.
\item $D$, the set of \emph{divergent} values. Evaluating these values will fail, therefore neither this nor any subsequent clause will be matched.
\end{itemize}

The values in these sets are represented by triples of the form $(\Gamma, \overrightarrow{v}, \Delta)$, where:
\begin{itemize}
\item $\Gamma$ is a typing environment that keeps tracks of variables and type variables.

For variables, it is a map from variables that occur in $\overrightarrow{v}$ to types. We denote that $x$ has type $\tau$ by $\Gamma \vdash x : \tau$. 

For type variables, it simply records their existence in the context, written as $\Gamma \vdash a$.

Let $x\#\Gamma$ denote that a variable or a type variable $x$ does not occur in $\Gamma$.

\item $\overrightarrow{v}$ is a vector of patterns, where pattern can be
	\begin{itemize}
		\item A variable, as in \h{id x = x}.
		\item A guard $G = (P, e)$ where $P$ is a pattern and $e$ is a Boolean expression.
		\item A data constructor pattern $K \overrightarrow{p}$, for example \h{Just a} of \h{Maybe a} and $\overrightarrow{p}$ is a vector of patterns. Note that $\overrightarrow{p}$ may be empty, e.g. for \h{False}.
	\end{itemize}
\item $\Delta$ is a set of term and type equality  constrains. The can be of form
	\begin{itemize}
		\item $x = e$, where $x$ is a variable and $e$ is an expression; $e$ and $x$ may be of any Haskell type.
		\item $x = \bot$, where $x$ is a variable and $\bot$ represents a divergent computation.
		\item $\tau_1 \sim \tau_2$, a type equality.
	\end{itemize}	
\end{itemize}

The algorithm defines three functions $\mathcal{C}\overrightarrow{p} A$, $\mathcal{U}\overrightarrow{p} A$, $\mathcal{D}\overrightarrow{p} A$ that take a vector of patterns $\overrightarrow{p}$ and an abstraction $A$ and compute the set of covered, uncovered, and divergent value abstraction respectively.

The first iteration starts with the most general value abstraction, $U_0 = \left(\left\{(x_1, \tau_1), \dots, (x_m, \tau_m)\right\} , x_1 \dots x_m, \emptyset \right)$. For clause $i$, we compute the  abstractions from the fall-through values and filter out those that are not plausible:

\begin{align}
C_i = \left\{w \middle| v \in U_{i - 1}, w \in \mathcal{C} \overrightarrow{p_i} v, \vdash_{\textsc{SAT}} w \right\} \\
U_i = \left\{w \middle| v \in U_{i - 1}, w \in \mathcal{U} \overrightarrow{p_i} v, \vdash_{\textsc{SAT}} w \right\} \\
D_i = \left\{w \middle| v \in U_{i - 1}, w \in \mathcal{D} \overrightarrow{p_i} v, \vdash_{\textsc{SAT}} w \right\}
\end{align}

The $\vdash_{\textsc{SAT}}$ denotes satisfiability \emph{over-approximation} for set of constraints, i.e. $\forall k. \not\vdash_{\textsc{SAT}} k \Rightarrow\quad \not\vdash k$, provided by an \emph{oracle}, as discussed in section \ref{sec:algOracle}. The algorithm is independent of a particular oracle or its properties.

\subsection{Uncovered values}

We now describe $\mathcal{U}$ in more detail. Covered and divergent values are analogous. \Cref{tab:uncov} gives its definition as a Haskell-style function from a pattern vector and a single value abstraction that represents one possible input to a set of possibly uncovered abstraction refinements.

The \textsc{UNil} rule states that for an empty pattern vector and an empty value value abstraction vector, there are no uncovered values. This is only useful for constants and to terminate recursion.

The \textsc{UConCon} rule applies when both the pattern and the abstraction are constructor patterns. When the constructors are equal,\footnote{Observe that they must be constructors of the same type, otherwise the program is not well-typed.} then the arguments ($\overrightarrow{p}$ and $\overrightarrow{u}$) are extracted and the computation continues on the flattened list. Mapping \texttt{kcon} then reconstructs the structure by applying $K_i$ to the appropriate number of resulting abstractions.

In case the constructors do not match, the value abstraction is definitely uncovered because it will not be matched, so it is returned unchanged.

In contrast, \textsc{UConVar} is matching a constructor pattern against a variable value abstraction. To find the possible values of $x$ for which $K_i$ will not match, each possible constructor $K_j$ of the type is substituted and the constraints are recorded, and a union of recursive solutions is taken. For all $K_j \neq K_i$, the recursive computation will return a non-empty set due to \textsc{UConCon}. The recorded constraints can then filter out implausible abstractions.

\textsc{UVarVar} assumes equality of the two variables and computes the remainder recursively. This describes all the abstractions that are unmatched due to constructor inequality at a subsequent position.

The \textsc{UGuard} shows how the Boolean expression $e$ is added to the set of constraints and substituted by a fresh unique variable, which prevents aliasing of possibly unrelated values in the \acrshort{SMT} constraints.

\subsection{Oracle}\label{sec:algOracle}
An \emph{oracle} is a Boolean function $\vdash_{\textsc{SAT}}$ of value abstraction triples.
For a triple $(\Gamma, \overrightarrow{v}, \Delta)$, it serves to over-approximate whether the constraints in $\Delta$ are satisfiable.

Since $\Delta$ is consists of value equalities as well as type equalities, this can be implemented using a separate term-level constraint solver and a type-level constraint solver respectively, both of which must then over-approximate factual satisfiability.

In particular, a trivial oracle that declares any input to be satisfiable is sound with respect to the properties given in \Cref{sec:soundness} and only decreases precision.


Note that for non-\acrshortpl{GADT} or \acrshortpl{GADT} with trivial constraints, all type equalities are trivial, i.e. of form $a \sim a$, thus do not have any effect on precision.

\subsection{Recommendations}\label{sec:warnings}
Recommendations are generated from the results of the analysis as follows:
\begin{itemize}
\item If the uncovered value abstraction set $U_k$ for the last clause is non-empty, then all its value abstractions represent missing clauses.

\item For every clause $i$ that has an empty set of covered value abstractions $C_i$, there are no values that can be matched, thus the clause may be redundant. If the set of divergent value abstractions $D_i$ is also empty, then the clause $i$ is redundant, otherwise it has an inaccessible right-hand side.
\end{itemize}

\subsection{Evaluatedness}
After running the discussed analysis for a function with $k$ clauses, we obtain the analysis \emph{trace}, a list of 3-tuples 
\begin{align}
	\left[ (C_0, U_0, D_0),\ \dots,\ (C_k, U_k, D_k) \right]
\end{align} 
containing the values abstractions corresponding to the definition in \Cref{sec:algo_out} after each of the $k$ iterations. 
The trace is used to compute the evaluatedness of the function.
The value abstraction patterns in $D_i$ represent the divergent values of each of the arguments in clause $i$.
We then assert that each argument (or subexpression thereof) that gets a constraint of the form $x = \bot$ will be evaluated during pattern matching.
This corresponds to an evaluation of $\bot$.

The result of this post-processing is a trace of tuples of value abstraction vectors.
The first element in this tuple specifies the form of the input and the second indicates how and which arguments are evaluated.

For example, in the case of our running example \h{f}, the evaluatedness is as follows:
\begin{minted}{raw}
f a b        
a: _         
b: b         
\end{minted}

\begin{minted}{raw}
f a False
a:     a
False: False
\end{minted}

It is to be read as:
When \h{f} is evaluated with input of the form \h{f a b}, where \h{a} and \h{b} are not further specified, only \h{b}'s first constructor is evaluated during pattern matching.
Further, when \h{f} is evaluated with input of the form \h{a False}, \h{a} will be evaluated.

Not all divergent values need necessarily cause the pattern matching evaluation to diverge.
Consider the following function that takes an arbitrary tuple as an argument:

\begin{minted}{haskell}
fst :: (a, b) -> a
fst (x, _) = x
\end{minted}
The evaluatedness of this function is simple:

\begin{minted}{raw}
fst a
a: a
\end{minted}
It states that the first argument to \h{fst} is always evaluated to its first constructor.
Indeed, when \h{fst undefined} is evaluated, the evaluation diverges during pattern matching.
It is, however, not evaluated beyond the first constructor, which in this case is the tuple constructor \h{(,)}.
Conversely, when \h{fst (undefined, undefined)} is evaluated, the tuple is matched but the arguments are not further evaluated and therefore the evaluation does not diverge during pattern matching.

\subsection{Soundness properties}\label{sec:soundness}
The over-approximation of satisfiability results in the following properties:
\begin{itemize}
	\item If there are no value abstractions in $U$, then there exist no concrete values that are not covered.
    In other words, the non-exhaustiveness warnings are sound.
	\item Similarly, $C$ is an over-approximation.
    Therefore any clauses reported as redundant indeed are redundant.
	\item By the same token, reported inaccessible RHSs are indeed inaccessible.
\end{itemize}

\subsection{Complexity}
Algorithms for manipulating patterns are known to be predominantly exponential. For example, determining the set or redundant clauses has been shown to be NP-complete \cite{sekar1992adaptive}. 

It is easy to see that the algorithm runs in $O(n\overline{m}c^{\overline{m}})$ time where $n$ is number of clauses, $\overline{m}$ is the maximum number of patterns occurring in any clause, and $c$ is the maximum number of constructors of any data type occurring among parameters.

This is because \textsc{UConVar} of \Cref{tab:uncov} establishes the upper bound on the number of value abstractions at any given point (maximum $c$-fold increase relative to the input abstraction) and the abstraction size is constant in $\bar{m}$. This also implies space usage of $O(n\overline{m}c^{\overline{m}})$ and $O(nc^{\overline{m}})$ satisfiability queries of size $\overline{m}$.\footnote{Assuming the constraints are solved incrementally, as outlined in \cite[Section 6.2]{DBLP:conf/icfp/KarachaliasSVJ15}.}

\section{Implementation}
Our implementation is publicly available\footnote{\url{https://github.com/PJK/haskell-pattern-matching}} under an open source license. In this section, we discuss some additional technical considerations; \Cref{sec:evaluation} showcases the tool in practice.

\subsection{Overview}
Each function is analyzed separately.
The clauses, guards, and other relevant information is extracted directly from the function's \acrlong{AST}; no interaction with the \acrshort{GHC} interface is required. Minimal desugaring is performed~(see \Cref{sec:desugar}).

This \acrshort{AST} passed to the main analysis algorithm, which produces the analysis trace, which is not yet filtered using $\vdash_{\textsc{SAT}}$ at this point. Each value abstraction in the trace is fed into the oracle (\Cref{sec:oracle}) to query satisfiability of the recorded constraints. This gives the final analysis result that contains plausible value abstractions only. All the warnings are generated from the filtered trace; the initial \acrshort{AST} is used to augment the output so that it closely resembles the input source code (\Cref{sec:warnings}). 

Since \acrshort{GHC} interfaces are not required for the structural analysis or the guard exhaustiveness issues we focus on, we avoid the laborious integration with \acrshort{GHC} altogether. This entails some limitations on the language features we support. In particular, without an access to the \acrshort{GHC} type-constraint solver \cite{DBLP:conf/icfp/SchrijversJSV09}, we can only provide a rudimentary \acrshort{GADT} support.

\subsection{Desugaring and special types}\label{sec:desugar}
Throughout preceding sections, we have assumed that all types are defined uniformly using standard definition of constructors. Since our tool operates predominantly on the \acrshort{AST} level, built-in types and values that require special syntactical support have to be addressed before the analysis.

For lists, we simply define the empty list constructor 
\h{[] :: a -> [a]} 
and the infix concatenation constructor 
\h{(:) :: a -> [a] -> [a]}
A list of the form 
\h{[x, y, ..., z]} 
is then translated into 
\h{x:(y:(...:(z:[]))}

In the same manner, tuples are defined as using a single constructor 
\hc{(, ... ,) :: a -> ... -> z -> (a, ..., z) }
So as to increase clarity, both tuples and lists are translated into their original syntactic forms before output.

Integers and other numerals are also a challenge, since they conceptually have an infinite number of constructors.\footnote{Haskell integers are implemented using GMP arbitrary precision arithmetic. Conceptually, however, \h{data Integer =  ... | -1 | 0 | 1 | ...} is a valid data type.}
This limitation is overcome by replacing integer literals with a variable pattern and a guard pattern that asserts equality. For example, \h{g 42 = 1} would be translated to (e.g. \h{g x | x == 42 = 1}).

Wildcard patterns (\h{f _}) and user-provided guards are also subject to desugaring, as outlined by Karachalias \cite[Figure 7]{DBLP:conf/icfp/KarachaliasSVJ15}.

Generating correct \acrshort{SMT} formulae also involves adding reconciliation of type definitions. For example, values of numeric types such as \h{Word8} must be postulated to be within their range in order to capture this property within the formula. 

\subsection{Oracle}\label{sec:oracle}

Term-constraints, in our simplified context, can only be variable equalities: $v_1 = v_2$, bottom assertions: $v = \bot$ or Boolean equalities: $v = e$, where $e$ is any Boolean Haskell expression
The Boolean equalities come from guards in the function under analysis.
\footnote{See the UGuard part of \Cref{tab:uncov}.}

\subsubsection{Translating expressions}
When querying the oracle, the set of constraints is checked for satisfiability.
Boolean expressions as they appear in Haskell, however, code cannot simply be fed to a satisfiability solver.

An expression is first broken down into a simple abstract syntax tree.
This tree is then translated into a SMT representation.
There is support for certain functions like \h{&&}, \h{+}, \h{not}, and other Boolean and numerical functions that are supported by the solver, but not all Haskell expressions can be fully translated.

For example, a Boolean function like \h{isPrime :: Int -> Bool} is not broken down at all.
Instead, the sub-expression \h{isPrime x} is treated as a Boolean variable by itself, because it can be either True or False depending only on \h{x}.

\subsubsection{Resolution of term-constraints}
To judge satisfiability of term-constraints, first all variable constraints $v_1 = v_2$ are resolved by replacing the variable $v_1$ by $v_2$ in all other constraints.

The next step handles all sets of term constraints with bottom assertions.
Any bottom assertions $v = \bot$ makes the set of constraints unsatisfiable if the variable $v$ occurs in another constraint. 
This means we can entirely discard a set of term constraints as unsatisfiable when we find such a bottom assertion.

Once the bottom assertions are dealt with, the only expressions that are left are Boolean equalities.
These are then passed directly into Z3 theorem prover~\cite{DBLP:conf/tacas/MouraB08} via the SVB library~\cite{sbv}.

\subsubsection{SMT results}
The results from this solver are then converted into an over-approximation of satisfiability.
Any unsatisfiability is interpreted as such, but any other result, whether it be ``satisfiable'', ``timeout'' or ``unknown'', is interpreted as ``satisfiable''.
Finally, the unsatisfiable value abstractions are removed from the analysis traces to complete the analysis.

\section{Evaluation}\label{sec:evaluation}

As explained in a previous section, our implementation does not depend on the sophisticated \acrshort{GHC} infrastructure, focusing on the term equalities instead. Since we also omit most language extensions and thus are unable to process most real-world code bases, we perform the evaluation on a qualitative basis.

For functions with non-GADT data types and no guards, our implementation gives exactly the same results as the \acrshort{GHC} version 8.0.1 (released May 21, 2016) with \verb|-Wall| flag, which we use as baseline for all subsequent comparisons. This already constitutes a major improvement over the \acrshort{GHC}~7 implementation \cite[Section 7, Table 1]{DBLP:conf/icfp/KarachaliasSVJ15}.

For functions with guards, we see an improvement in precisions, i.e. see warnings that \acrshort{GHC} misses due to its coarse $\vdash_{\textsc{SAT}}$ over-approximation.

\subsection{Integer constraints}

Consider the following (erroneous) implementation of the absolute value function:

\begin{minted}{haskell}
abs :: Int -> Int
abs x
    | x < 0 = - x
    | x > 0 = x
\end{minted}
The code compiles without any warnings, even though \h{abs 0} is undefined. Our tool indeed does detect the non-exhaustive definition and provides a counterexample:
\begin{verbatim}
The patterns may not be exhaustive, the following 
clauses are missing:
abs x
[...]
x = 0 :: Integer
\end{verbatim}

\subsection{Boolean constraints}
In the same manner, our tool also improves precision for Boolean guards. Consider the following example of a redundant guard:
\begin{minted}{haskell}
bguard :: Bool -> Int
bguard x
    | x         = 0
    | not x     = 1
    | otherwise = 2 -- redundant
\end{minted}
Even though $\neg x \wedge \neg (\neg x)$ is unsatisfiable, the \acrshort{GHC} solver cannot reveal the inconsistency (it can only discover inconsistencies of the form $x \wedge \neg x$), thus failing to report the redundancy, whereas our tool reports the following recommendation.

\begin{minted}{raw}
The following clause is redundant:
bguard x | otherwise
\end{minted}

\subsection{Mixed constraints}

Apart from from improving the precision for integers and Booleans, our approximation of other Haskell fragments can improve precision when guards contain e.g. function applications.

For example, the following function guards contain an unknown function \h{isPrime}:
\begin{minted}{haskell}
isPrimeAndSmall :: Int -> Bool
isPrimeAndSmall x
    | isPrime x && x < 10 = True
    | not (isPrime x)     = False
\end{minted}
Nevertheless, we can still show that the definition is not exhaustive by treating \h{isPrime x} as a symbolic expression and give a counterexample:
\begin{minted}{raw}
The patterns may not be exhaustive, the 
following clauses are missing:
isPrimeAndSmall ~a
Constraints:
  ~f == False
  ~f == not (isPrime x)
  ~c == False
  ~c == isPrime x  ~a < 10

Satisfiable. Model:
[...]
\end{minted}

\subsection{Future work}
A surprising but straightfoward applications of our work lies in increasing the typechecking precision in languages that enforce totality constraints in a sound but incomplete way. For example, consider the following Rust \cite{DBLP:conf/sigada/MatsakisK14} implementation of the signun function:

\begin{minted}{rust}
fn sgn(x: i32) -> i32 {
  match x {
    y if y < 0 => -1,
    y if y == 0 => 0,
    y if y > 0 => 1,
  }
}
\end{minted}

The Rust compiler\footnote{As of version 1.23.} will refuse this function as the pattern is possibly incomplete. Using the analysis techniques we present, such imprecisions can be eliminated up to the level afforded by the oracle. Similarly, the value-level constrains we generate could also be useful during program optimization as it is reasonable to expect that they are more precise than commonly used dataflow analyses.

The next step in extending the analysis of constraints would be to also fully process guards that are defined in terms of functions and other data types. 

For functions, it remains unclear whether re-formulation in e.g. uninterpreted functions theory is feasible. In particular, all functions used in the constraints would have to be total, which cannot be enforced in Haskell as of now, but the area is a subject of active research \cite{liquid}.

Proving properties of general data structures is, within a limited scope, viable. Zeno \cite{DBLP:conf/tacas/SonnexDE12} and HipSpec \cite{DBLP:conf/cade/ClaessenJRS12} have demonstrated implementation of the concept for Haskell, but both are no longer maintained and do not support the current language ecosystem.

Finally, with the shift towards dependent typing, many properties will become a part of the type system and a significant portion of the work might thus be offloaded to the type-level constraints solver.

\acks
The authors would like to thank Tom Schrijvers for the supplementary material to his paper as well as his encouragement.

\clearpage
\Urlmuskip=0mu plus 1mu\relax
\sloppy
\bibliographystyle{IEEEtran}
\bibliography{report}


\end{document}